# Multiple Aging Mechanisms in Ferroelectric Deuterated Potassium Dihydrogen Phosphate

Gregory A. Fields, Samuel F. Cieszynski, Bo Zhao, Kidan A. Tadesse, Mohammed A. Sheikh, Eugene V. Colla, and M. B. Weissman

*Department of Physics, University of Illinois at Urbana-Champaign*

*1110 West Green Street, Urbana, IL 61801-3080, U.S.A.*

## ABSTRACT

The anomalously large dielectric aging in ferroelectric partially deuterated potassium dihydrogen phosphate (DKDP) is found to have multiple distinct mechanisms. Two components cause decreases in dielectric response over a limited range of fields around the aging field. Substantial aging occurs on time scales of ~1000s after a field change, as expected for a hydrogen/deuterium diffusion mechanism. A slower component can give almost complete loss of domain-wall dielectric response at the aging field after weeks of aging. There is also a particularly unusual aging in which the dielectric response increases with time after rapid cooling.

**<u>Introduction</u>**

The high-dielectric response plateau regime of the ferroelectric (FE) phase of $KH_2PO_4$ (KDP)[1], and partially deuterated $KD_{2x}H_{2(1-x)}PO_4$ ($D_XKDP$)[2] [3], has large dielectric susceptibility arising from a dense array of parallel planar 180° domain walls[3-5] $D_XKDP$ has been reported to show surprisingly large reduction in this dielectric response while aging, e.g. ~40% reduction in response over 22 hours aging slightly below the Curie temperature, $T_C$.[6] In contrast, KDP itself shows less than ~2% aging slightly below $T_C$.[6-8]

The large aging effect in $D_XKDP$ has been attributed primarily to diffusion of H to domain walls.[6] This redistribution of H is thought to occur because the increase of $T_C$ with deuteration[2,9-11] implies thermodynamically that free energy is lowered when an H is in a paraelectric region and D is in a ferroelectric one, compared to the reverse arrangement. Since domain walls lack full FE order, one suspects that in equilibrium in $D_XKDP$ sheets of higher H and lower D concentration should accumulate at the walls, although the change in free energy need from H accumulating at the walls need not be the same as that for H moving to a bulk paraelectric region.

On the other hand, a closely related ferroelectric, $RbH_2PO_4$ (RDP), also shows giant aging effects for which such diffusion mechanisms are not considered to be a major source,[8] raising the question of whether multiple types of major aging effects are present in $D_XKDP$. In this paper we show that the simple model of H diffusing to domain walls can account for the properties of only one component of the aging effect, and that at least two other major aging effects are also present. (For a fuller discussions of different types of aging, along with more extensive background references, found in general disordered systems see ref.[12] and for ferroelectric systems specifically see ref. [8].)

Here we show that there are very long-time aging components that can remove most of the domain-wall response at the aging field, as extreme as what has been seen in RDP[8], even for temperatures just a little below $T_C$. Although the very long-time results have only been taken at zero field, we show that aging for times of several days is approximately independent of dc field, but not of field history. We show that the long-time tail of the aging does not fit the kinetics expected for the simple diffusion mechanism, and that the kinetics of aging and of forgetting the aging effects after field changes differ for the long-time tail but not for the initial fast effect. The aging effects are found to be particularly huge for intermediate values of x. Finally, we show that under some conditions there is a highly anomalous upward aging, in which the dielectric response increases as a function of aging time.

Optical images of the domain structure show that the large aging cannot result from simple domain growth leading to loss of net domain-wall area, since that domain growth is not large.[6] Instead, the domain walls become less mobile due to increased pinning by disorder. Several key lines of evidence point to the conclusion that the large aging comes from changes in the underlying disorder that pins the domain walls, rather than just in domain walls gradually finding well-pinned states on fixed disorder. The most significant is that aging at a particular electric field leaves a persistent hole near that field in plots of susceptibility vs. field, i.e. reduced susceptibility only for fields near the aging field. [6]

Page 3                                                                4/21/19

Reduction of susceptibility by domain-wall pinning to fixed disorder would, in contrast, produce a hole that would not persist after field changes large enough to move the walls far from their old positions.[6] Some such persistence is, however, expected for alterations in the disorder, which would not immediately dissipate after the domain walls moved.

There is indication that H-D diffusion is a major mechanism for the change in underlying disorder.[6] First, there is the obvious contrast between $D_X$KDP and KDP, which shows little aging[6,][7,8], especially when the temperature is not close to the putative domain-wall-glass regime in which the domain-wall response is lost at low T.[4,7,13,14] Second, the aging in $D_X$KDP exhibits a characteristic time on the order of 100s, seen both in the dependence of second-harmonic generation on field-sweep-rate and in re-aging of susceptibility after small quakes in the domain configuration.[6] This time scale is about what is expected based on H-D mutual diffusion rates of ~$10^{-15}$ $cm^2$/s [15] and the domain-wall width of ~2.5 nm[16].

One possible reason, not previously explored, for large aging effects to be found in a commercial $D_X$KDP sample but not home-grown KDP samples would be that some sorts of disorder other than the H-D distribution might be present in the commercial material. In this paper we report initial aging results on home-grown samples with a range of x and report some key detailed data on several such samples, showing that key features are not unique to the commercial $D_X$KDP material but occur for almost all samples with x > 0.2. Particularly large aging is found for x= ~0.4. (We have also found two samples with

x=0.68 that show relatively little aging in the first few hours. We have not yet conducted detailed studies of these unusual samples.)

The simplest aging picture for $D_XKDP$ would involve homogeneous diffusion of H to replace D at domain walls, since these walls lack FE order. The strong increase of $T_C$ with x [2,9-11] implies that the free energy difference between FE and paraelectric phases is a decreasing function of x. A close analogy exists to known aging mechanisms in ferromagnets[17] and other ferroelectrics[18], in which point-like defects diffuse to domain walls. One can estimate the magnitude of the attraction of H to the domain walls if one makes the rough guess that the domain wall environment is similar to the PE state, since the dependence of $T_C$ on x is known ($dT_C/dx = \sim 107K$) [2,9-11], as is the latent heat of the first-order PE-FE transition[2,19,20] over a wide range of x (E.g. at x=0.8 the latent heat is about 317 J/mole[19]). The approximate value of the free-energy reduction as a single H (out of the two per chemical unit) in an FE region exchanges location with a D from a PE region for material with $T_C$ of 211K is then given by thermodynamics to be $\sim(107/211)*317J/2N_A = 1.3*10^{-22}$ J, where $N_A$ is Avogadro's number. That energy is less than a tenth of the thermal energy, $k_BT$ (where $k_B$ is Boltzmann's constant and T is temperature), in the temperature range of interest. Thus if we assume that the free energy change of an H-D exchange between an FE region and a domain wall is similar to that of an exchange between an FE region and a PE region, one would expect that the equilibrium formation of an H-rich sheet at the domain walls would nearly follow a linearized Boltzmann distribution, and that the kinetics would be well approximated by a simple linearized diffusion process. Most importantly for our current experiments, one would expect that if the domain walls were moved to new positions via an applied electric field, the sheet formed at the original domain wall position on aging at the original field would diffuse away following the same kinetics with which it formed. We shall see that this similarity of aging and forgetting kinetics only holds for the initial part of the aging effect.

We shall describe another surprising effect: upward aging. After rapid cooling the dielectric response grows over time rather than reduces, in sharp contrast to aging effects in a wide variety of disordered materials.[12] Optical images of the domain structure in the rapidly cooled material indicate that the upward aging comes from the gradual formation of the parallel-sheet structure from a patchy initial pattern .

Finally we shall discuss possible implications of the aging effects for the principle question remaining for KDP and related ferroelectrics- the origin of the rather abrupt freezing transition by which domain walls lose their mobility at low temperature. [3,4,13,21]

Model Background

Since a key point of this paper will be that the linearized diffusion model does not suffice to describe the full aging process, before presenting results we shall describe the kinetics expected for such a model. A straightforward calculation of the kinetics of such a model can be done in spatial Fourier space, since each spatial Fourier component of the H concentration decays exponentially in time toward its Boltzmann equilibrium value, with time constant inversely proportional to the square of the wavevector. By approximating the shape of the domain-wall potential well as a Gaussian, one obtains an easily solvable analytic model in which H accumulating at domain walls gives an excess binding free energy of the wall that is proportional to $(1 - \left(1 + \frac{t}{\tau}\right)^{-\frac{1}{2}})$, where t is the elapsed time since the H distribution was uniform and $\tau$ is a characteristic diffusion time, roughly the square of the domain wall width divided by the diffusion constant. The $t^{-1/2}$ tail of this form is insensitive to the shape assumed for the domain wall potential, simply reflecting the form of the diffusion process in one-dimension. This form would be appropriate to model dielectric response consisting of jumps between discrete free-energy-minima domain-wall positions, with the free-energy depth of any minimum in which a domain wall sits increasing with H accumulation. If one makes the approximation that the reduction of the susceptibility is proportional to the increased pinning potential, one obtains a $t^{-1/2}$ tail:

Eq. (1) $\quad \varepsilon(t) = \varepsilon(\infty) + \left(\varepsilon(0) - \varepsilon(\infty)\right)\left(1 + \frac{t}{\tau}\right)^{-\frac{1}{2}}$.

For response driven by simple field-tilting of a quadratic domain wall free-energy minimum, the linear susceptibility would fall off inversely with the *curvature* of the pinning potential, just like the response of a spring to a force. The H-binding contribution to the curvature would increase as $(1 - \left(1 + \frac{t}{\tau}\right)^{-\frac{3}{2}})$, from the same analytical calculation, again with the $t^{-3/2}$ form of the tail depending only on the diffusion equation, not the exact shape of the domain-wall potential. The reason for the different exponent is that the curvature is dominated by higher spatial wavevector components, which form more quickly. Given that there is a large out-of-phase response even in

Page 7 4/21/19

the linear regime, we believe that the discrete-response model, hopping between minima, is more appropriate than the steepening quadratic well model.

In any such a model, application of a dc field large enough to move domain walls out of the potential wells formed by H accumulation resets the aging process, although the form of the recovery will depend on how far the domain walls have moved from the H-rich sheet. Application of ac fields large enough to create non-linear response will lead to broader, shallower H-rich sheets reflecting the average Boltzmann factor over the oscillating domain-wall positions, and thus give reduced aging. That qualitative effect, however, is shared with other disorder-rearrangement aging mechanisms.

**Materials and Methods**

Highly deuterated DKDP crystals were obtained from United Crystals. Other $D_X$KDP crystals were grown by us from water/heavy water solutions, by gradually reducing the solution temperature. We grew $D_X$KDP crystals by cooling KDP-DKDP solutions, prepared to be saturated at 80°C, starting from 90°C at 1K per hour with no seed. We checked the value of x using our measured $T_C$'s. We use x = ($T_C$ - 122K)/107 K ±0.02 for a calibration curve, a consensus reached in several careful studies in which x of the crystal itself was measured independently, using several different techniques.[9-11] Since the commercial crystals had $T_C$=211 K, they actually had x=0.83 ±0.02, despite their nominally higher specified x of 0.95. (The supplier confirmed that 0.83 is a plausible value.)

Samples were made into capacitors with the c-axis orthogonal to the contact surfaces by cutting with a diamond saw and lightly polishing. Contacts were made by depositing a thin layer of Cr (~10nm) followed by about 100nm of Ag via thermal evaporation. Sample thicknesses ranged from 0.7 mm to 1.2 mm, with contact areas of ~ 6.5 mm$^2$. Optical samples were polished using lapping discs with grit size down to 0.3 µm. Deposited electrodes for the optical samples were offset laterally ~1mm to allow unobstructed view of a region between the electrodes, so that for these samples the applied fields were not parallel to the c-axis. Digitized optical images of the domain structure, including movies taken after field changes, were obtained via polarized light microscopy using a Leica DM2700 microscope.

Our dielectric response measurements were made via a standard lock-in amplifier with applied ac voltage at 100Hz, along with adjustable dc voltage up to ±100 V in one set-up or ±400 V in another. The dielectric response of KDP and D$_X$KDP is highly non-linear even at low fields, due to weakly pinned domain walls. [14] Since some of the interesting features of the response are most evident in the non-linear regime,[14] most of our measurements were made with 1 V rms ac voltages, corresponding to 8-16 V/cm ac fields. Some runs were made with 100 mV ac voltages (0.8-1.6 V/cm), close to the linear response regime, and others at 5V ac voltages (40-80 V/cm), well into the non-linear regime. We also ran some checks of the aging at much lower ac fields, finding qualitatively similar results. For brevity we will refer to the dimensionless polarization/field ratio as ε', although ε' formally should only refer to the low-field limit of that ratio. We approximately removed a baseline due to background stray capacitance, measured by fitting $\varepsilon'(T)$ well above $T_C$ to a Curie-Weiss law.

Several experimental protocols allowed separating the effects of aging at different fields. In one, after cooling at E=0 and then aging at a field of one sign the field was switched to opposite sign, in order to move the domain walls from the positions at which they had already had a chance to age. Then as the sample aged at the new field brief several-second excursions were made to the original field, to track how the response there forgot the effects of the prior aging. In a related protocol, after aging at an initial field the field was steadily swept over a range including the aging field, tracking both how the response at the initial field forgot its aging and how the response at a range of other fields aged. In a combined protocol, the sweeps were begun only after the measurements of multiple excursions back to the initial aging field. In still another protocol, the magnitude of the applied ac field was switched between large and small values.

In some protocols, we cooled the sample very rapidly (>10K/min), with the initial intent to follow a shorter time scale of initial aging. We found instead qualitatively different behavior, *upward* aging of the response, and tracked its time course after stopping cooling at several different temperatures in the plateau regime.

## Results

Fig. 1 shows the in-phase and out-of-phase dielectric response for samples of two materials described in this paper. The results are very similar to those previously reported on materials of this sort, e.g. [14]

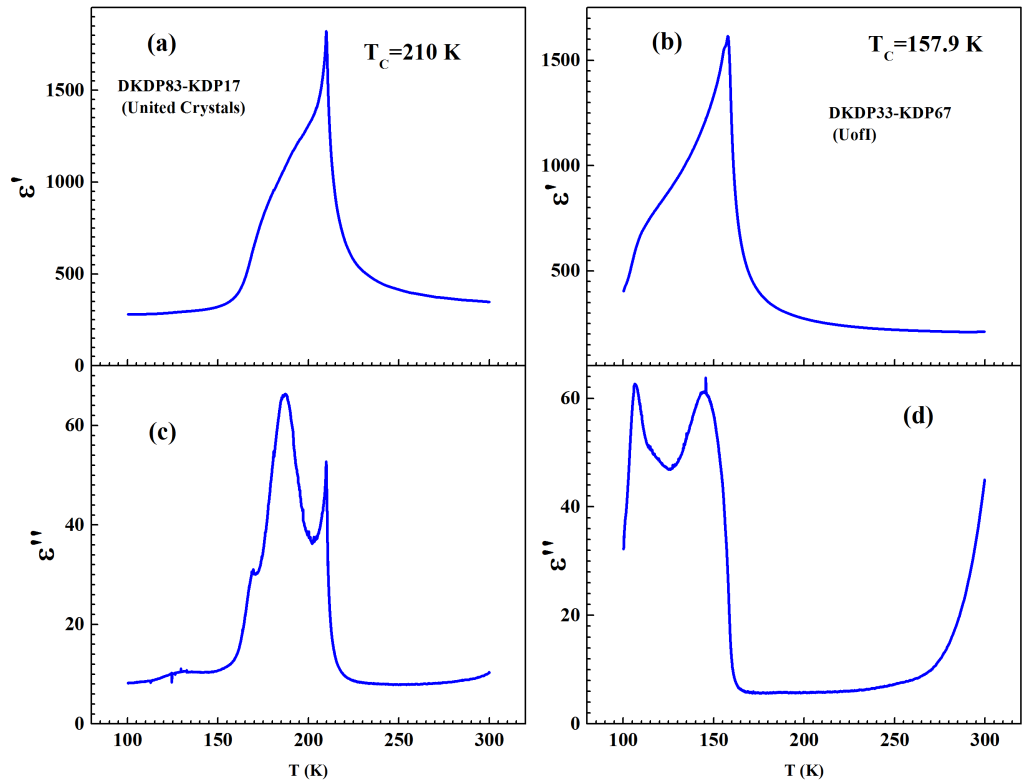

Fig. 1. The in-phase and out-of phase response at 100 HZ with an applied ac voltage of 1V rms is shown as a function of temperature for a commercial sample with x=0.83 (panels (a) and (c), taken at 9.3 V/cm rms) and a home-grown sample with x= 0.33 (panels(b) and (d), taken at 8.2 V/cm rms).

Fig. 2 shows two stills from a digital movie of the domain structure in a commercial sample taken at 193 K after cooling through $T_C$ at ~4K/min. Several important results are apparent. The domain structure consists of parallel sheets, as expected. These sheets are not highly regular, however, but show variations in thickness. As the sample ages the number of domain walls shows negligible decrease on the time scale of ~ 1 hour over which the susceptibility decreases by ~ 20%. In the movie, one can see that occasional localized rearrangements of the domain structure occur for several hours. These images, clearer than the ones in our previous paper,[6] confirm that simple domain growth is not a significant source of aging, but they do indicate that local changes in domain configuration can continue for a long time.

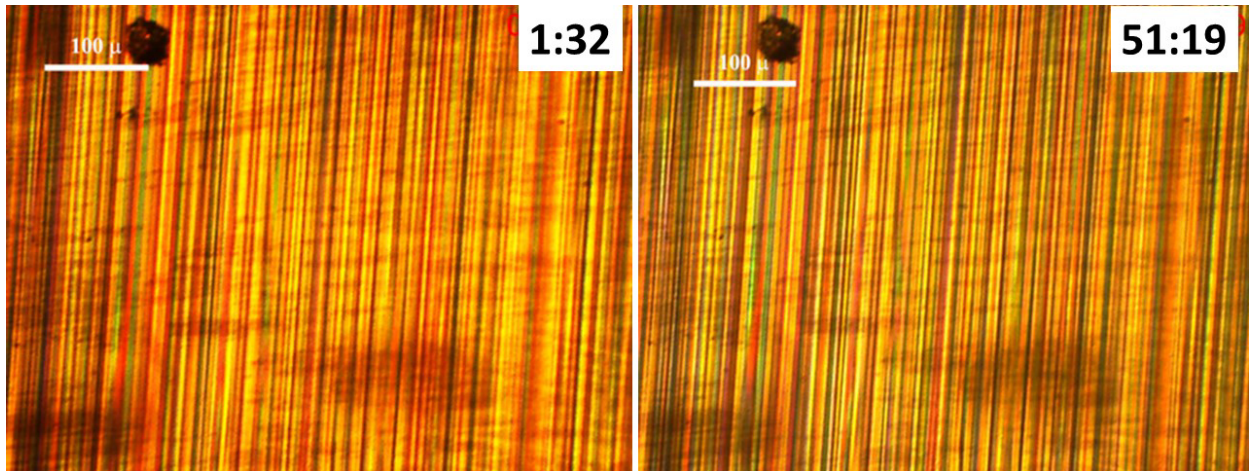

Fig. 2 shows two stills from a digital movie of the domain structure in a commercial x=0.83 sample taken at 193 K, at about 1.5 min and 51 min after cooling through $T_C$ at ~4K/min.

Our previous results[6] described aging of the effective susceptibility taken with 1V rms ac voltages, giving ac fields of roughly 10V/cm rms. Since the non-linear susceptibility and the linear susceptibility can show qualitatively different behavior [14], we checked whether the aging of the linear susceptibility measured at small ac fields resembled that measured at large fields. We also checked whether the very large aging was unique to the commercial material or would be similar to that of material grown in our lab. As shown in Fig. 3, very large aging was also found in the home-grown material. The kinetics at different ac fields were qualitatively similar but not quantitatively identical. Slower aging is found at the largest ac fields, >30 V/cm. At lower fields, the aging rate shows small apparently non-monotonic dependence on ac field, but this is within the range of run-to-run variability. Thus most of our aging results, taken at larger fields, should be taken to give only the qualitative behavior of the linear response. The ag

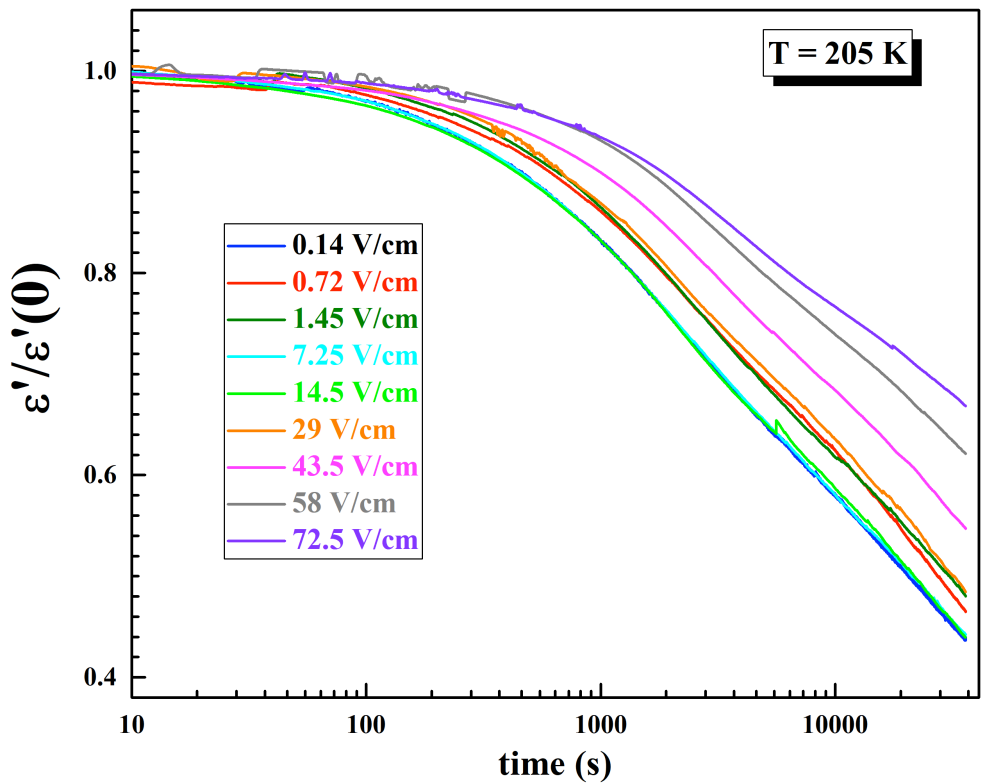

Fig. 3 shows aging taken at a range of ac fields after cooling at 4K/min from above $T_C$. The lowest ac field is within the range of almost completely linear response at this temperature. [14] The sample is home-grown with x=0.83.

In order to check the functional form of the loss of susceptibility we measured the aging over very long times. Fig. 4 shows very long-time aging of ε' taken at T=205K over 11 days for a homegrown sample with x=0.83. Similar results, with an even more prominent long-time tail of the decay, were found on a commercial sample, with one 11-day run at 200K and, with a still more prominent tail, on a 15.8 day run using a dry-ice bath at ~195K rather than a cryostat. The long-time tail of the aging was surprisingly large, with the small-signal ε' dropping to under 15% of its initial value after 11 days.

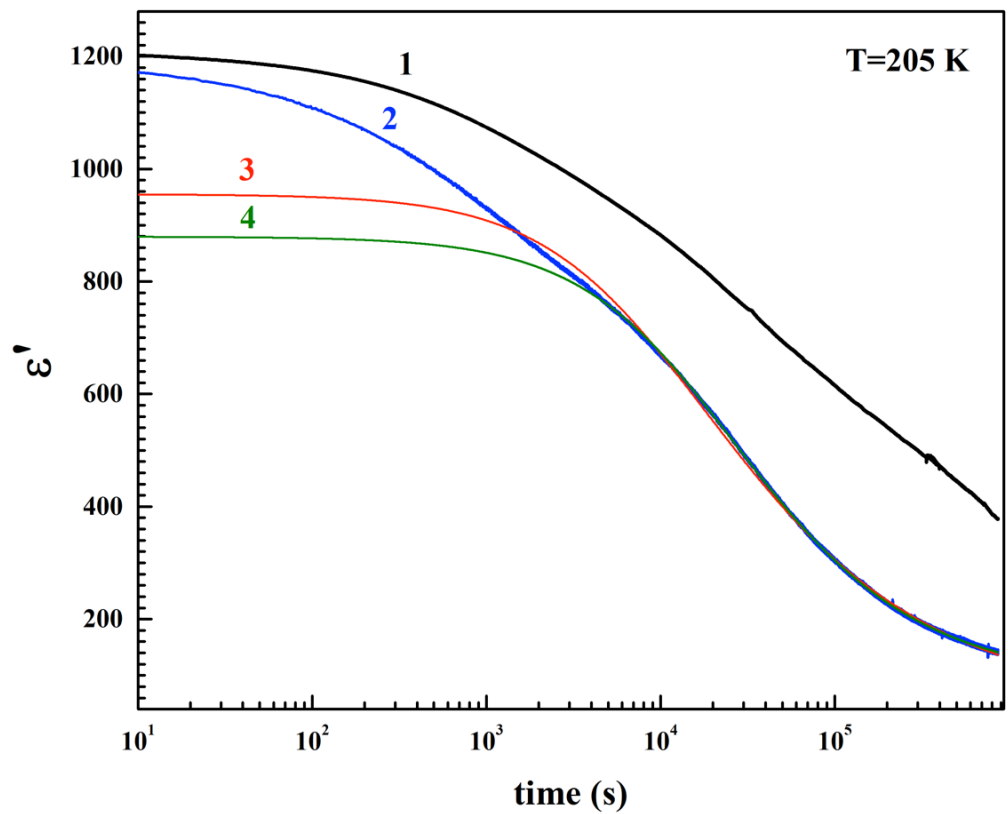

Fig. 4. ε'(t) is shown for a home-grown x=0.83 sample at 205 K after a dc field switch of 1450 V/cm. Curve 1 shows results with an rms ac field of 72.5 V/cm and curve 2 shows results with a field of 1.45 V/cm, near the linear range at this temperature. Curve (3) shows fitting of the long-time tail (t > 3000s) of the 1.45 V/cm ac ε'(t) aging to Eq. 1. A better fit (curve 4) is obtained with an exponent of 0.7 rather than 0.5, but still fails to capture the short-time behavior.

The tail of the 1.45 V/cm ac data have far too large a long-time tail to fit the $t^{-3/2}$ form expected for simply increasing the curvature of the potential wells in which the domain walls sit. As is clear on Fig. 4, the shape of the long-time tail depends on ac voltage. Due to the length of time of each run, we have not yet systematically explored this dependence. Comparison with data on other samples and other temperatures shows that the form is not universal, although the tail is always much larger than the $t^{-3/2}$ form of the simplest diffusion picture.

Although the tail of the 1.45 V/cm data in Fig. 4 can be approximately fit via Equation (1), which assumes that the effect on ε' is linear in the binding potential, the fit requires using τ = ~9000s and thus misses most of the initial decay, as shown. The best fit with an adjustable exponent, ~0.7, requires an even longer τ = ~19,000 s, and also fails to fit the initial decay.

We've found that reasonably simple functional forms that fit the long-time tail leave a positive short-time residual and likewise fits to the short-time portion leave a positive long-time residual. That suggests the possibility that these components have different origins. We shall see via more complicated experimental protocols that the initial component is qualitatively different from the slower part, with only the initial decay (< ~1000 s) showing the reversibility expected for the H diffusion effect.

All these aging effects change ε' only over a range of E around the aging field[6], in contrast to what would be found for domain growth. Fig. 5 shows very broad field sweeps taken immediately after aging a commercial sample for 1 hour at 200 K. As the field is swept, a hole in ε'(E) ( measured at 1 V rms ac) emerges near the field at which the sample was aged. All the field-sweep data show a consistent hysteretic effect, with the hole appearing at fields slight more positive than the aging field on positive-going sweeps and at fields slightly more negative than the aging field on negative-going sweeps. The hysteresis effect is much larger than would be produced by the lock-in response time and thus reflects the physics of the domain wall pinning. The hysteresis is expected if the domain wall positions do not follow the equilibrium position determined by the field, but lag behind their equilibrium positions due to partial pinning by disorder.[8,22]

At fields ~ 3000 V/cm away from the aging field, the effects on ε' become small. Qualitatively similar behavior was found in a sample with x=0.46 at T=155 K, but with the width of the hole reduced by about a factor of four.

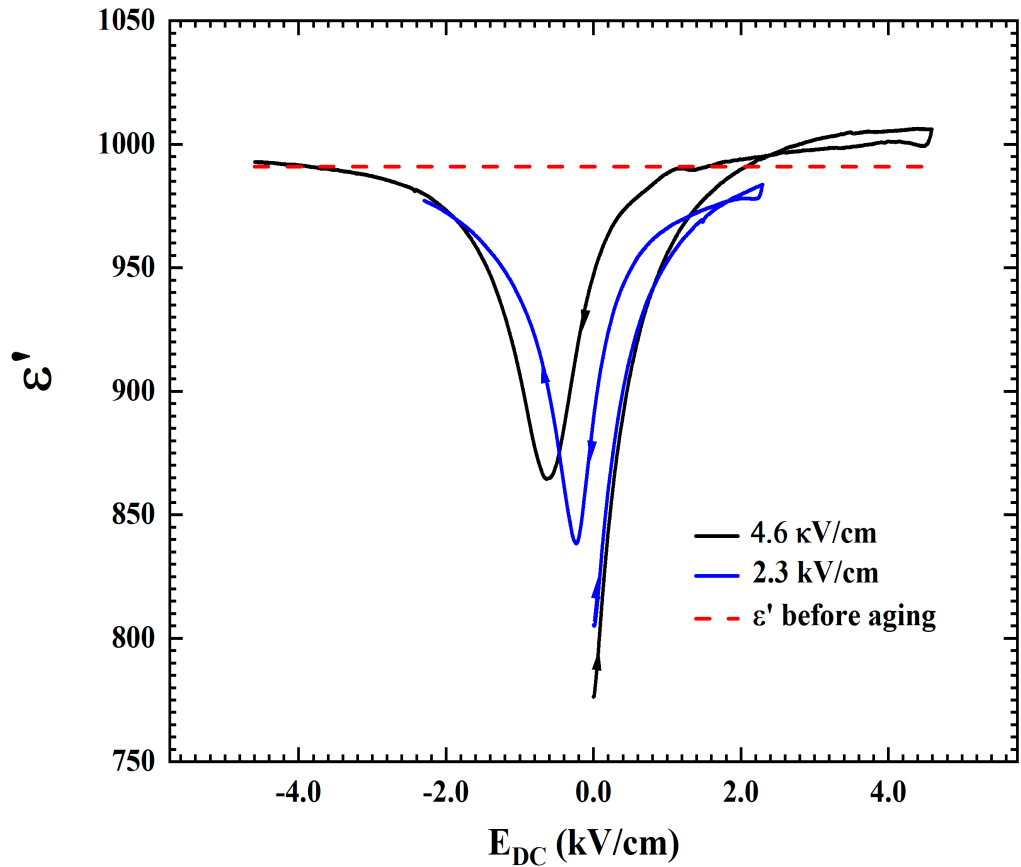

Fig. 5. ε' vs. E taken at 200K after aging a commercial x=0.83 sample at E=0 for 1 hour. The sweep rate was ~ 3 V/cm-s. Arrows show sweep directions. The dashed line shows ε'(0) before aging, approximately equal to ε'(E) over this range.

Since the long-time tail of the aging kinetics appears incompatible with a simple H-D diffusion model, supporting the likelihood of multiple aging mechanisms, we looked for other possible difference in the kinetics or field dependence of the short-time and long-time components. We ran a series of experiments to check not only the kinetics of the initial aging but also the kinetics with which it was forgotten when the field was changed.

To check the time evolution of ε' (E,t) over a range of E, we used continuous field sweeps over ±1140 V/cm after initial aging for 11 days at 200K and E=0 V/cm. The sweeping took place over the course of 15 hours with each individual scan taking ~6 minutes. A representative sampling of the field-increasing part of the scans is shown in Fig. 6, with early time scans sampled more densely to exhibit the more rapid short time behavior. During the continuous sweep ε' continues to decrease for fields out in the wings of the initial hole, even as it increases near the original aging field, i.e. forgets the aging effects.

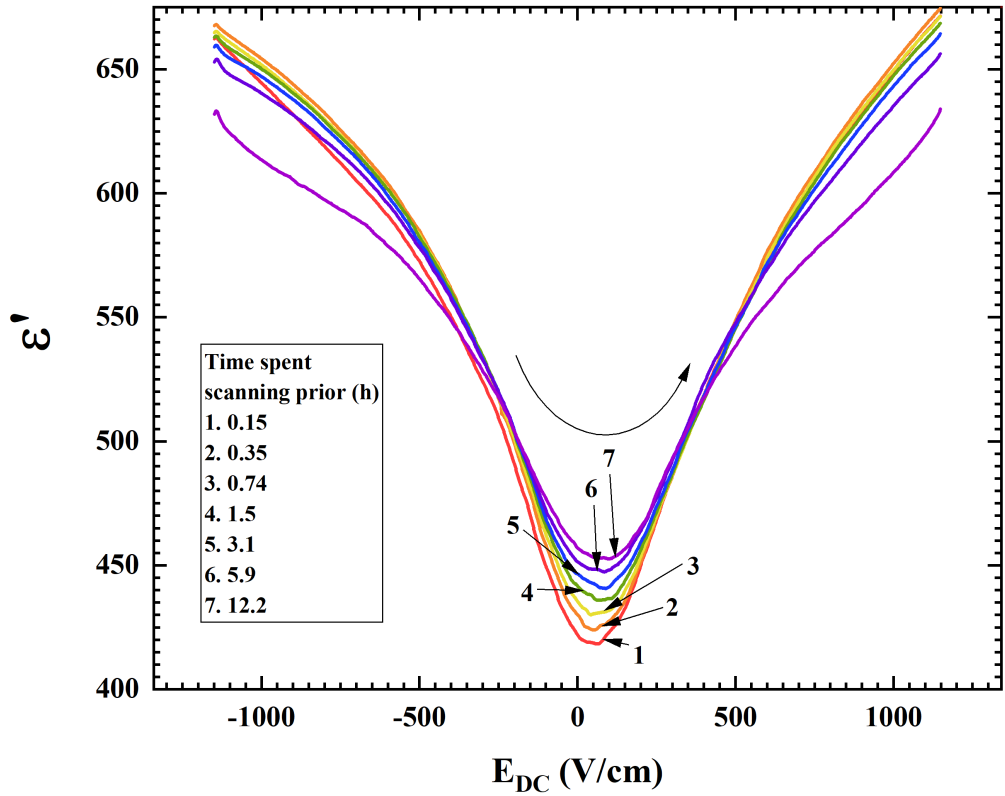

Fig. 6 shows a sampling of the field-increasing scans (as indicated by arrow) following an 11 day aging experiment at T= 200 K on a commercial x=0.83 sample. Here each scan took 6 minutes. The field-decreasing sweeps are similar to the field-increasing ones shown, but with the ε' (E) holes shifted to negative E rather than positive E.

Figure 7 illustrates that time dependence of ε' in the hole, partway up the wings, and at the edge of the sweep range.

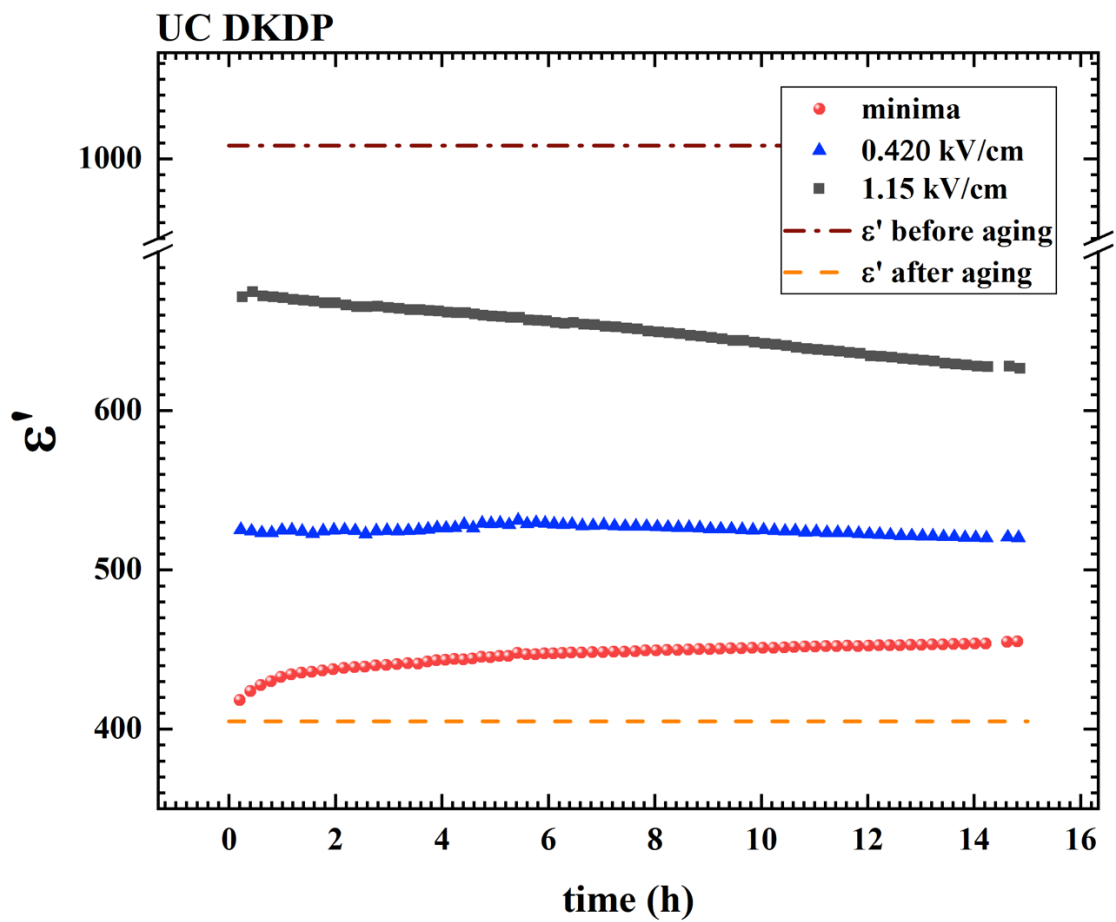

Fig 7. ε'(t) during the sweeps shown in Fig. 6 is shown at the minimum value, at 420 V/cm, and at 1150 V/cm.

The general conclusion from the sweep experiments is that the initial aged hole is lost during the field sweeps, while an additional less sharply field-dependent aging forms. The results fit a picture in which the aging of ε'(E) is the sum of a term that reflects the short-time aging from time recently spent with field near E and a term that represents how much time the field has spent near E on the long time scale.

One conclusion emerges directly from the sweep data. The results seem incompatible with a simple model in which the domain walls are modeled as rigid objects described by a one-dimensional position, with uniform response to electric fields, and with aging simply due to steepening potential wells from H accumulation at the wall position. In such a model, a field application sufficient to move the walls by several wall thicknesses would pull the walls out of the wells and eliminate the aging effects. The tails of the hole, such as those in Fig. 6, extend somewhat beyond 1000 V/cm from the aging field for the x=0.83 samples. The polarization change for a ~1000 V/cm field corresponds to ~1% of the saturation polarization, corresponding to domain wall motion of ~0.5% of the domain-wall spacing. Based on the optical images, that would be ~20 nm displacement, which is more than four times the domain-wall thickness measured by X-ray methods.[16] A picture of simple domain-wall motion within potential wells for a 1-D position fails in another even more basic regard. The effect would depend on the second derivative of the potential, which for reasonable potentials would have an integral of zero over the large sweep range. In most of these data there are at most very small positive tails to the negative hole, so the aging effect does not come close to integrating to zero over this range. Together with the prominent hysteretic effect on the position of the minimum in $\varepsilon'$, these results point toward a model in which flexible domain walls that move by hopping between local minima develop some more strongly pinned regions, i.e especially deep minima, [7,8] allowing partial memory of the aging to persist for larger field excursions.

Clarifying the kinetics by which the aging at the initial field is lost is complicated in sweep experiments because of the application of a broad range of fields in a sweep experiment. To simplify the interpretation of the forgetting kinetics, we aged the sample at one field, then switched to another field. Brief excursions (9s each) back to the original field then allowed tracking the kinetics with which the aging effects at that original field were forgotten, complicated only by the superposition of small aging effects from the single distinct new field.

In the experiment illustrated in Fig. 8, we aged a commercial sample at E= -2174V/cm for 21 hours at T=202K, then switched the field to E=+2174V/cm. Fig. 8 shows $\varepsilon$' (t) at both the new field and the original one. Qualitatively similar behavior was found after shorter initial aging times and also on an x=0.46 sample.

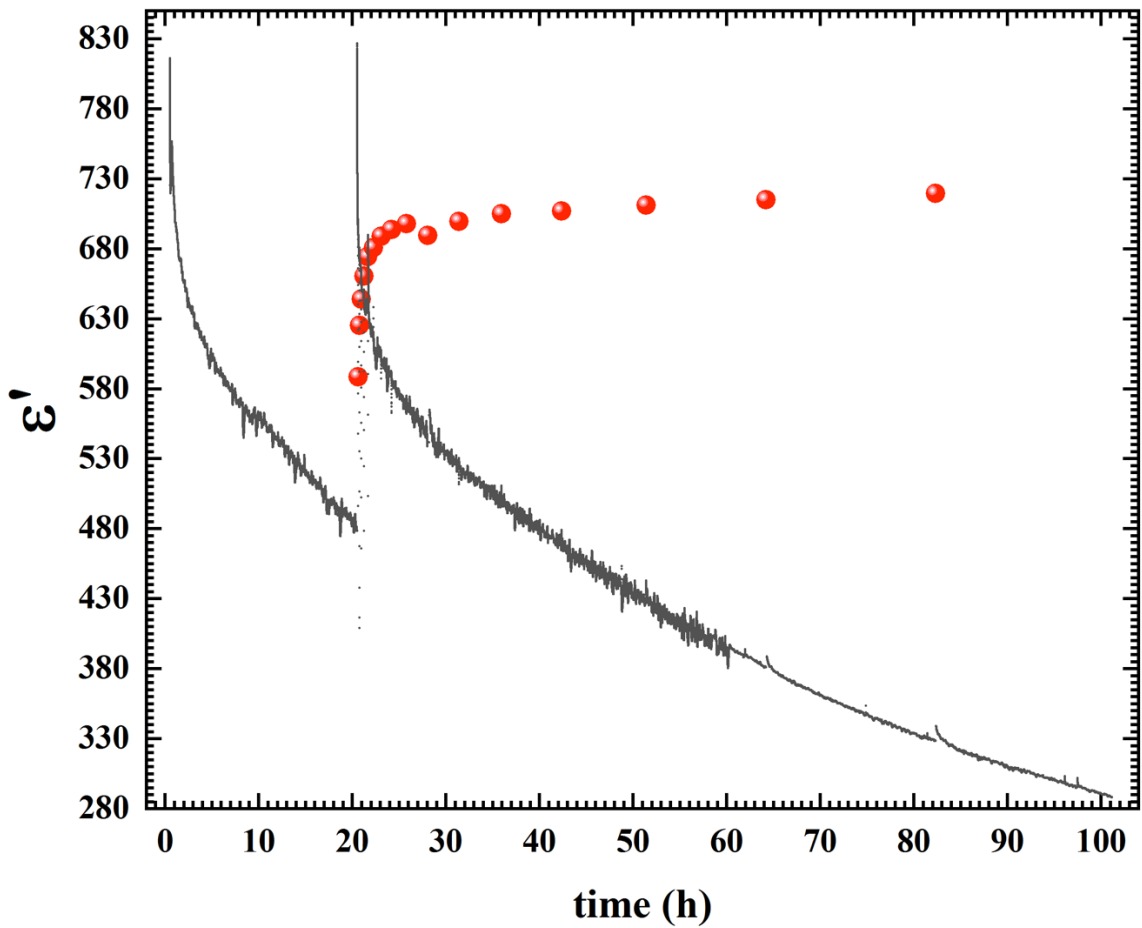

Fig. 8 shows the aging at the new voltage (=+2174V/cm) and initial 21-hour aging at the original voltage (-2174V/cm) as a function of time. The red dots show the recovery of the susceptibility at the original field as measured in 9 second excursions to the original field.

The aging at the new field shows no new features, indicating that the occasional brief switches back to the original field are not causing some sort of qualitative disruption of the domain structure. The forgetting kinetics at the old field are quite revealing. In the first hour after the field switch, ε' at the original field increases by approximately the same amount that ε' at the new field decreases. This fast component of the effect is thus reversible, as would be expected for simple linear diffusion kinetics.   Over the next 80 hours, however, ε' at the original field barely increases at all, while ε' at the new field decreases by more than a factor of two. ε' at the original field does not return to its initial pre-aged value.  The new field is so far from the old field that the tail of the aging hole at the new field cannot account for any significant fraction of that discrepancy. Thus the long-time aging does not show the reversibility of the short-time aging, a qualitative distinction.

A detailed look at the initial aging kinetics at a range of temperatures confirms the distinction between the initial fast aging and the subsequent slow aging, as illustrated in Fig. 9. For times of 4 s < t < 300 s after a field jump, the slope of the aging curves shows very little temperature dependence in the range from 175K to 190K, for which the domain wall width is nearly constant.[16]  A slight temperature dependence of the slope is found in that time range at higher T. For 1000 s < t < 20,000 s, in contrast, the aging rate varies by more than an order of magnitude over this temperature range. That contrast is consistent with H tunneling being the dominant kinetic mechanism for the initial aging, with some thermally activated processes being dominant at longer times.

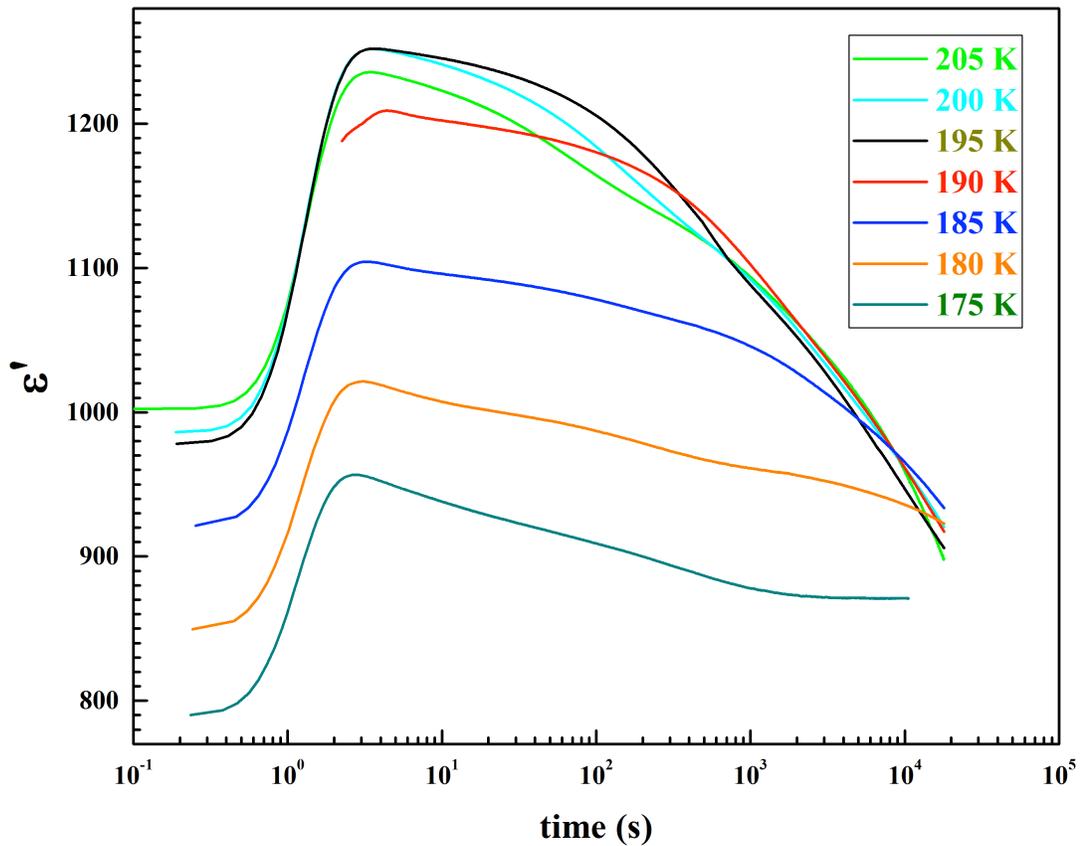

Fig. 9 shows aging of a commercial x=0.83 sample after field jumps of ~500 V/cm at a range of temperatures.

In pure KDP, we also found that the much smaller aging still had distinct fast and slow components, with qualitative results similar to those shown in Fig. 8. Since the natural abundance of D is very low, that indicates that H-D diffusion is not the only contributor to the fast aging in $D_X$KDP, although it appears to be the largest one.

Since aging in different ac fields follows somewhat different time courses, we checked whether that's simply due to the different mix of linear and non-linear responses measured at different fields or whether it's due to an effect of the ac field on the aging process itself. Fig. 10 shows the results of experiments in which the field was switched between 1.45 V/cm and 72.5 V/cm after one hour and two hours, and vice versa. The higher field is large enough not to be negligible compared to the width of the aging hole in ε'(E).

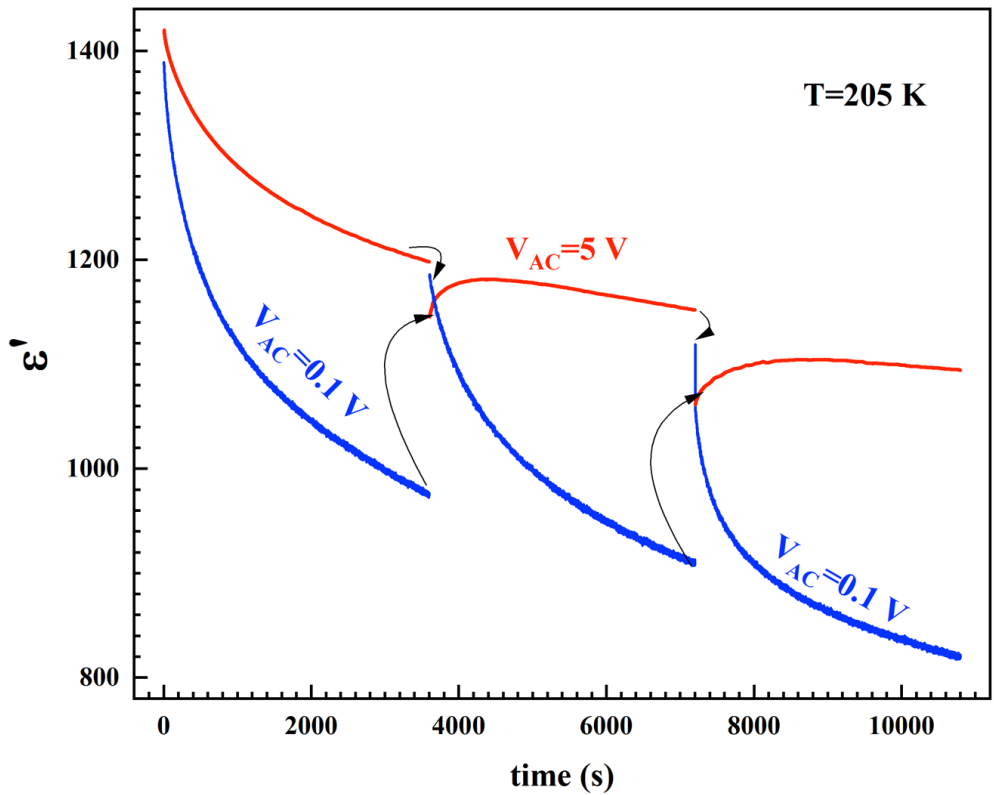

Fig. 10. Aging of ε'(t) of a homegrown x=0.83 sample of thickness 0.69mm is shown at two different ac voltages, 0.1V and 5V, i.e. rms fields of 1.45 and 7.25 V/cm. The protocols were to switch the dc field at t=0 to start fresh aging, run at one ac voltage for an hour, switch to the other for an hour, then switch back. Transients of ~6 s are omitted after each switch. Arrows show which data followed which.

After switching to the higher ac field, the high-field response increases, about 4%, with a characteristic time of ~200 s, in addition to the usual slow decrease. After switching to the lower ac field, a fast decrease of roughly 15% in the low-field response, also with characteristic time of very roughly 200 s, adds to the usual slow decrease. The slow decrease is also somewhat affected by the ac field, occurring more rapidly at the low field, as was also shown in Figs. 3 and 4. These results again indicate a qualitative difference between the short-time and long-time aging. One expects that the simple H accumulation mechanism would be reduced when the time-average domain wall potential is reduced in depth and spread out by the large ac, with some of the change immediate because the strongly driven wall sees less of the H sheet and some of the reduction appearing gradually as the H spreads out to the new time-average potential. The short-time behavior is thus in at least qualitative agreement with the expectations for the H-accumulation mechanism.

Large aging has not been found in KDP [6], so the pattern of the dependence on deuteration (x) may shed light on the mechanism. Although we have not yet systematically studied a wide range of x in our home-grown samples, we found large aging in all samples with $x > 0.2$ but not in samples with smaller x. The aging was particularly huge for samples with $x = 0.3$-$0.5$, as illustrated in Fig. 11.

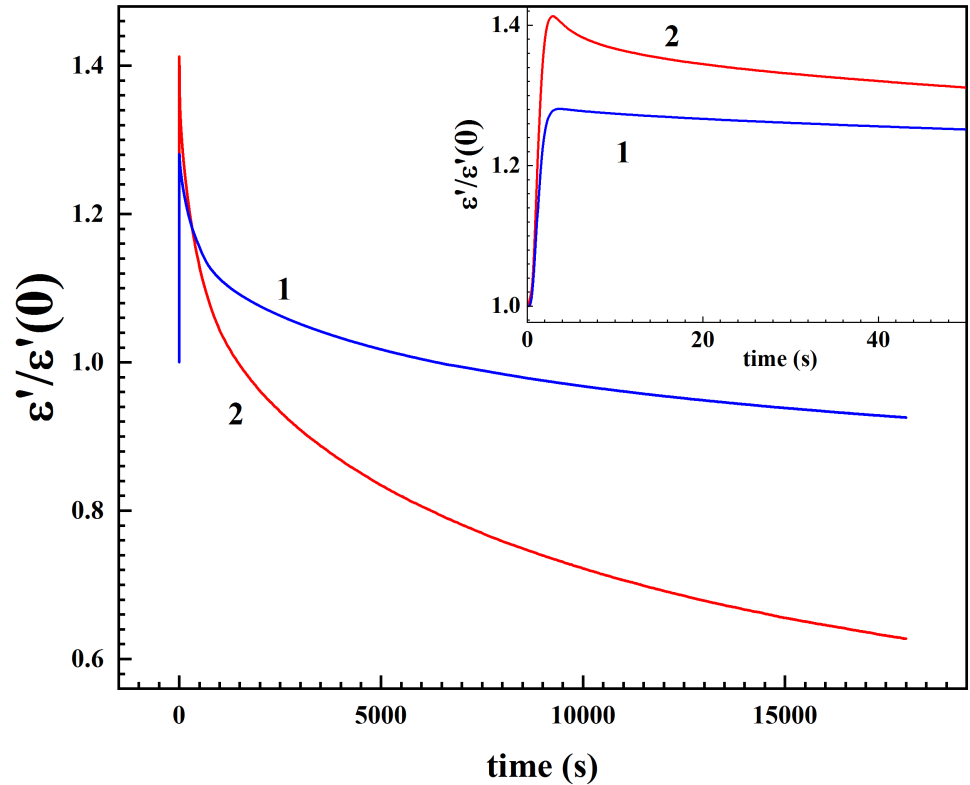

Fig. 11 Aging of ε' is shown for a commercial x=0.83 sample with thickness 0.60 mm (1) and a home-grown x=0.33 sample (2) with thickness 1.22 mm at temperatures of 0.92 $T_c$ and 0.9 $T_c$ respectively. Each sample was cooled in zero field to the aging temperature and then the dc field of about 350V/cm was applied. ε'(0) here refers to the value before the field change. The inset shows the initial evolution of the dielectric susceptibility.

In some protocols the samples were cooled rapidly (>10K/min) through $T_C$ to the target temperature. As shown in Fig. 12 for target temperatures in the range 170K to 195K for a commercial sample ε' increased initially after reaching the target temperature. At lower or higher temperatures the effect was not noticeable. From the time course of the upward aging at intermediate temperatures, it is clear that it has a strongly T-dependent rate. Above the range where upward aging is evident, it simply seems to be too fast to see, and below that range it is too slow to see.

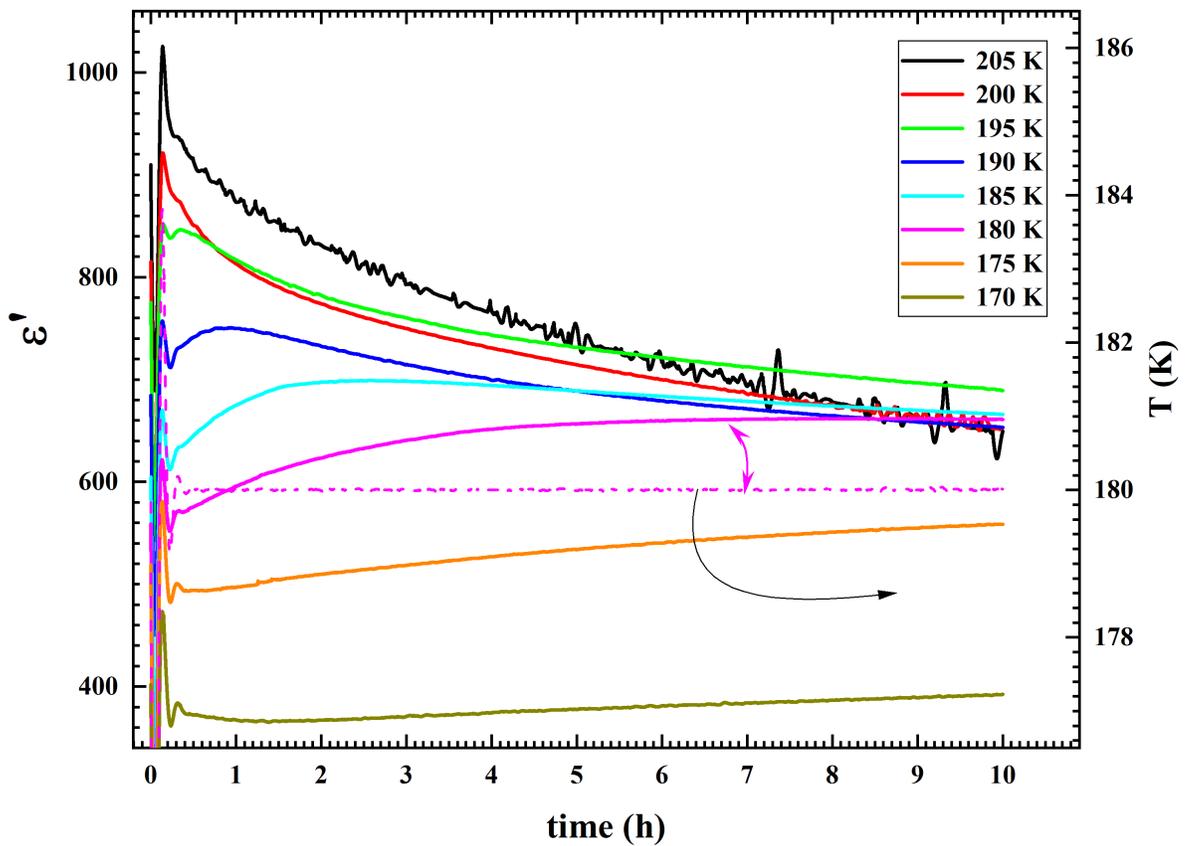

Fig. 12. Aging behavior at a range of temperatures of a commercial x=0.83 sample after rapid cooling, exhibiting upward aging on strongly T-dependent times. T oscillates for times under 0.5 hr, but is very stable after that, as shown by the dotted T(t) trace for the 180K data.

After the temperature stabilized ε'(t) could be fit fairly well with the product of the aging ε'(t) after slow cooling to the same T times a constant minus an exponential decay. In the range of 170 K - 185 K, the measured aging after slow cooling could be used in the fit. In the range 180 K- 190 K a fit to the fast-cooling result alone, but including an exponential fit to the falling ε'(t) rather than the slow-cooling data, worked well without too much uncertainty in the characteristic times of the rising time constant. The rise-times can be estimated well enough in the temperature range 175K to 190K to see that they fit an Arrhenius-like temperature dependence, shown in Fig. 13, with an attempt rate of about $10^{12.5\pm2}$ Hz, typical for local thermally activated processes, and an activation energy of 6800±800 $k_B$K, with generous error bars based on estimated systematic uncertainties in the fits. The data outside that range could also be fit with the same Arrhenius dependence, but with too much uncertainty in the characteristic times to be useful in determining the parameters.

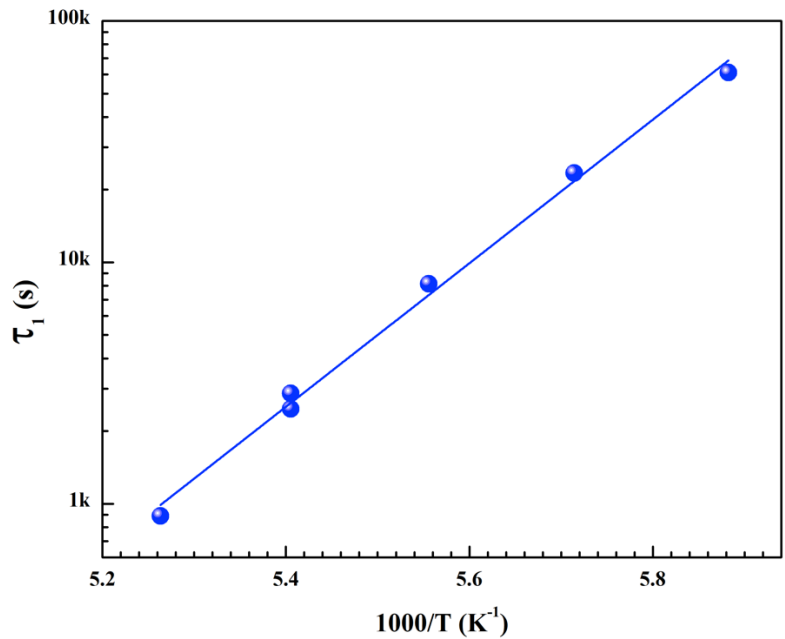

Fig. 13. The rise times for the upward aging are shown as a function of temperature for a commercial x=0.83 sample, along with an Arrhenius fit. The two points at T= 185 K (1000/T= 5.4 K$^{-1}$) are obtained by two different fitting procedures, giving an estimate of the systematic uncertainties. The same two procedures give almost identical results at T= 180 K.

The domain images taken after rapid cooling show many patches of domain sheets rotated 90° from each other, i.e. some along the a-axis and some along the b-axis, as seen in Fig. 14. Unlike the ordinary downward aging, the upward aging involves the annealing of these patchy domains.

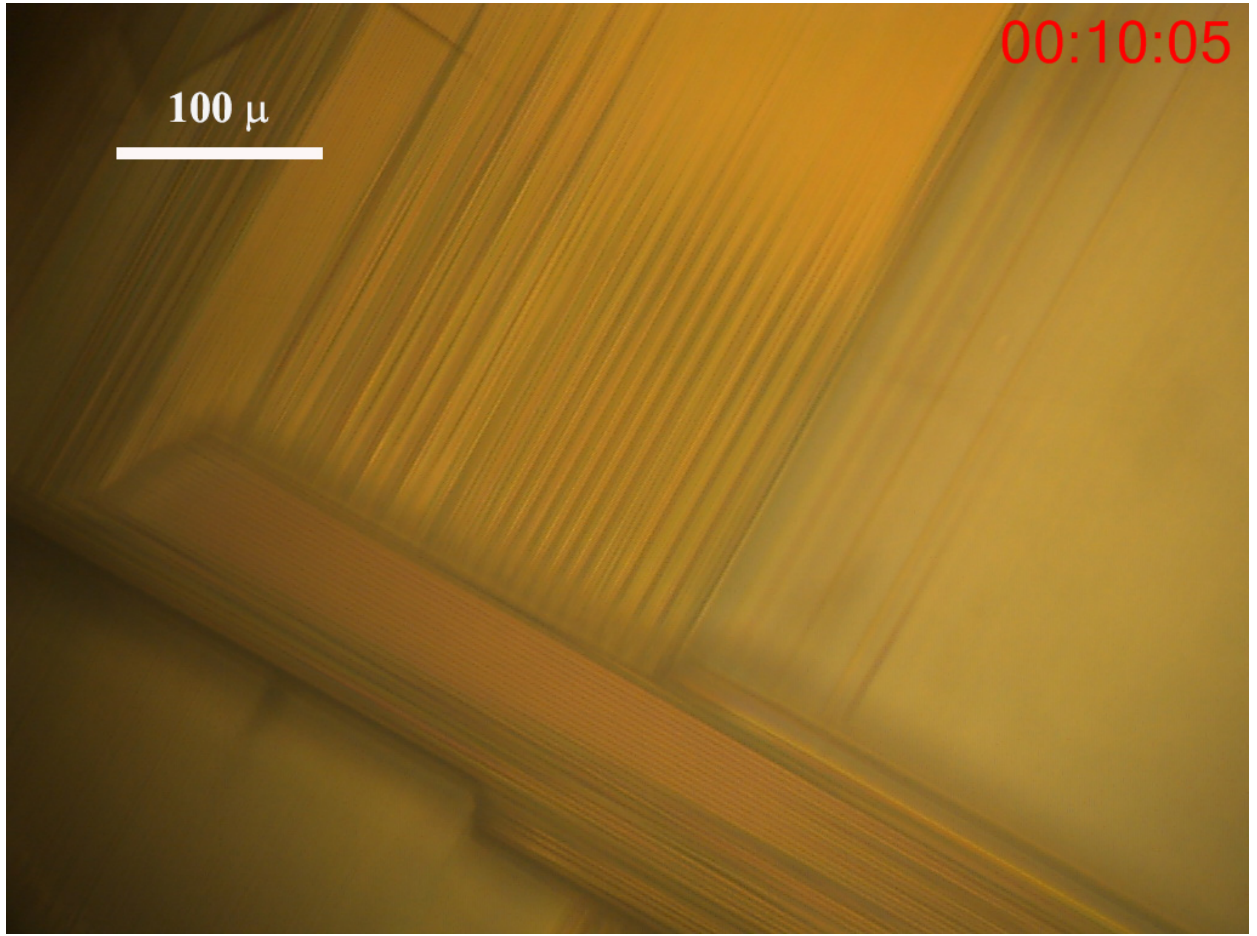

Fig. 14. A patchy domain pattern formed after rapid cooling (~30K/min) of a commercial sample through the transition to roughly 130 K.

**Discussion**

Perhaps the least expected result here is the upward aging found after fast cooling. Its mechanism appears not to be closely related to the other effects. This patchwork domain structure appears to reduce the domain-wall dielectric response, compared to that obtained from large parallel sheets. Initial annealing after quick cooling is dominated by the growth of the parallel-sheet structure. It's rate-limiting step involves a thermally activated process with a typical attempt rate ($\sim 10^{12}$ s$^{-1}$) for local motions. This novel upward aging appears to have little connection with the other aging mechanisms.

The most striking feature of the downward aging is that over periods of many days at temperatures a little below $T_C$, the dielectric response falls to almost to the same level as the response in the low temperature regime with frozen domain walls. RDP may show about as large an effect, although it has not been followed for quite as long.[8] Although there have been several ideas for why the domain-wall response is lost at the low-T end of the plateau region[3,4,13,21,22], none would imply that a similar loss of response would occur over time at a much higher temperature.

This downward aging gives holes in the ε' vs. E plots, with the hole centered near the aging field. That result confirms that modification in the underlying pinning potential is required, not just adjustment of domain walls to a fixed potential. [6,8] Since the aging in the linear response regime at very low applied ac fields is qualitatively similar to that in higher ac fields, and even a bit faster, this aging can occur via ordinary equilibrium kinetics, e.g. thermal diffusion, not requiring the ac driving.

The downward aging itself consists of at least two distinct components. One major component of downward aging remains consistent with an H-D diffusion effect giving an enhancement of H along the domain walls. This component of the hole in ε'(E) mostly forms on times shorter than ~1000 s and then is forgotten on a roughly similar time scale after the field is changed (see Fig. 8). Smearing out the time-averaged positions of the domain walls by application of a large ac field causes this component to be partially forgotten on that same time scale (see Fig. 10). Reduction of the ac field causes it to reform, again on about the same time scale. Its kinetics are not strongly temperature dependent, consistent with quantum tunneling (see Fig. 9), not surprising for H-D diffusion.

A second component of the downward aging also gives a hole in ε'(E) when E is held fixed, but with time scales of formation extending out to many days. Long-term aging at any field value lowers subsequent response at that field regardless of aging at other fields. Such an effect cannot be modeled by a simple domain-wall binding potential as a function of a one-dimensional variable representing domain-wall position since one cannot simultaneously have holes in the potential as a function of position at all positions in such a model. The results point toward a different type of model, in which local free-energy minima among which segments of the domain walls hop deepen with time.

Unlike the faster component, however, the forgetting kinetics and the aging kinetics for the slower component are not similar (see Fig. 8). Some of the long-time aging does not recover at all on the time scale of our experiments. An effect of this sort could be expected for defect rearrangements[8] that involve both large activation barriers and large free-energy differences between the aged state and the initial state. That combination can lead to relatively fast forward kinetics but slow reverse kinetics.

What are these defect rearrangements with large free-energy differences? Although the very large downward aging effects close to $T_C$ in $D_XKDP$ seem to require deuterium, the very large aging in RDP does not[8]. Field-history experiments on RDP are needed to see if its aging shares the qualitative features of the slow aging of $D_XKDP$. If it does, the likely suspect would be some structural rearrangement of disorder at the domain walls that occurs for only some range of parameters, shared by RDP and higher-x $D_XKDP$.

It is also possible that the reduced aging in low-x $D_XKDP$ simply arises because its ferroelectric phase is found at lower T, for which the activated arrangements are too slow. The extremely large aging for x of about 0.35 to 0.5, even compared to the x=~0.83 materials does not fit well with that idea. An intriguing but perhaps accidental coincidence is that a prominent anomaly in the entropy loss on forming the FE phase (including both the latent heat and some component from the heat capacity) has been reported at about the same x.[2] That raises the (perhaps remote) possibility that the domain walls are binding to distinct H-D ordered patterns that form for some x in the FE state, and that an unrelated effect is found in RDP.

Several follow-up experiments should help sort out the possibilities. It would be interesting to see whether the complicated field history effects found here are also present in RDP. A more systematic study of $D_X KDP$ materials with a range of x, including rare samples with reduced aging, may clarify what conditions are needed for the huge long-time aging.

Regardless of the detailed mechanism, we have seen that the binding of the domain walls to disorder grows very substantially during the aging of $D_X KDP$ just below $T_C$. Nevertheless, the temperature at which the domain walls freeze (the bottom of the plateau) is very weakly sensitive to that aging. That supports the picture of the domain-wall freezing as a sort of collective glassy crossover[13], for which the temperature scale is set by the stiffness of the domain walls and the interaction potential between them more than by the strength of the pinning potential.

**Acknowledgments**. S. F. Cieszynski was supported by NSF REU PHY-16-59598. G. A. Field's work was supported in part by a Philip J. and Betty M. Anthony Undergraduate Research Award from the Department of Physics at the University of Illinois at Urbana-Champaign. K. A. Tadesse was also supported by the Department of Physics.